\documentclass[twocolumn]{aastex61}
\usepackage{graphicx}
\usepackage{epsfig}
\usepackage{natbib}
\usepackage{multirow}
\usepackage{amsmath}

\newcommand{\eg}{{\rm e.g.,\ }}
\newcommand{\cf}{{\rm cf.,\ }}
\newcommand{\etal}{{\rm et al.\thinspace}}

\newcommand{\km}{{\rm\thinspace km}}

\newcommand{\s}{{\rm\thinspace s}}

\newcommand{\kmps}{\hbox{$\km\s^{-1}\,$}}

\newcommand{\halpha}{H$\alpha$}
\newcommand{\lya}{Ly$\alpha$}
\newcommand{\nhi}{$N_{\rm HI}$}

\newcommand{\hi}{H\thinspace{\sc i}}

\newcommand{\oiii}{[O\thinspace{\sc iii}]}

\newcommand{\oii}{[O\thinspace{\sc ii}]}
\newcommand{\oi}{O\thinspace{\sc i}}
\newcommand{\oivp}{O\thinspace{\sc iv}]}

\newcommand{\sivp}{S\thinspace{\sc iv}]}

\newcommand{\hei}{He\thinspace{\sc i}}
\newcommand{\sitwo}{Si\thinspace{\sc ii}}
\newcommand{\sithree}{Si\thinspace{\sc iii}}
\newcommand{\siiv}{Si\thinspace{\sc iv}}
\newcommand{\cii}{C\thinspace{\sc ii}}

\newcommand{\nv}{N\thinspace{\sc v}}
\newcommand{\fesclyc}{$f_{\rm esc}^{\rm LyC}$}
\newcommand{\fesclya}{$f_{\rm esc}^{{\rm Ly}\alpha}$}
\newcommand{\sigsfr}{$\Sigma_{\rm SFR}$}
\newcommand{\vpeaks}{$\Delta v_{\rm Ly\alpha}$}
\newcommand{\vchar}{$v_{\rm char}$}
\newcommand{\vmax}{$v_{\rm max}$}

\newcommand{\Msol}{\hbox{\thinspace M$_{\sun}$}}

\shorttitle{GP Kinematics}

\begin{document}

\title{Kinematics and Optical Depth in the Green Peas: Suppressed Superwinds in Candidate LyC Emitters \footnote{Based on observations made with the NASA/ESA Hubble Space Telescope, obtained at the Space Telescope Science Institute, which is operated by the Association of Universities for Research in Astronomy, Inc., under NASA contract NAS 5-26555. These observations are associated with programs GO-14080, GO-13293, and GO-12928.}}

\author{Anne E. Jaskot}
\altaffiliation{Hubble Fellow}
\affiliation{Department of Astronomy, University of Massachusetts, Amherst, MA 01003, USA.}
\author{M. S. Oey}
\affiliation{Department of Astronomy, University of Michigan, Ann Arbor, MI 48109, USA.}
\author{Claudia Scarlata}
\affiliation{Minnesota Institute for Astrophysics, University of Minnesota, Minneapolis, MN 55455, USA.}
\author{Tara Dowd}
\affiliation{Department of Astronomy, University of Massachusetts, Amherst, MA 01003, USA.}

\begin{abstract}
By clearing neutral gas away from a young starburst, superwinds may regulate the escape of Lyman continuum (LyC) photons from star-forming galaxies. However, models predict that superwinds may not launch in the most extreme, compact starbursts. We explore the role of outflows in generating low optical depths in the Green Peas (GPs), the only known star-forming population with several confirmed and candidate LyC-leaking galaxies. With {\it Hubble Space Telescope} UV spectra of 25 low-redshift GPs, including new observations of 13 of the most highly ionized GPs, we compare the kinematics of UV absorption lines with indirect \hi\ optical depth diagnostics: \lya\ escape fraction, \lya\ peak separation, or low-ionization absorption line equivalent width. The data suggest that high ionization kinematics tracing superwind activity may correlate with low optical depth in some objects. However, the most extreme GPs, including many of the best candidate LyC emitters with weak low-ionization absorption and strong, narrow \lya\ profiles, show the {\it lowest} velocities. These results are consistent with models for suppressed superwinds, which suggests that outflows may not be the only cause of LyC escape from galaxies.

\end{abstract}

\section{Introduction}
\label{sec:intro}
Star-forming galaxies are likely the dominant source of the Lyman
continuum (LyC) photons that reionized the universe at $z>6$
\citep[\eg][]{robertson15}. However, because of the high \hi\ column
densities near young star-forming regions, LyC photons do not easily
escape their host galaxy. To overcome this challenge, a variety of
theoretical and observational studies invoke mechanical feedback,
particularly by supernovae (SNe), as a means of clearing out \hi\ gas
and enabling LyC escape. \citet{clarke02} show that above a critical
star formation rate (SFR), overlapping SN-driven bubbles create a
porous interstellar medium through which LyC photons can
travel. Similarly, radiation hydrodynamic simulations find that
SN-driven outflows in low-mass galaxies boost their LyC escape
fractions \citep[\eg][]{wise09, trebitsch17}, with LyC
emission peaking in a star-forming region after SNe expel
surrounding gas \citep{ma15}. Observations suggest that high SFR
surface densities (\sigsfr) and high gas velocities may correlate with low optical depths \citep{alexandroff15}.

However, theoretical predictions are converging on a different
behavior in the regime of extreme feedback.  For the
most massive, compact, super star clusters (SSCs), superwinds
should be inhibited by strong radiative cooling \citep{silich07,krause14,yadav17} due to high
gas densities.  The high ambient
density and pressure can also suppress the growth of individual
stellar wind bubbles, thereby preventing them from merging into a
superwind \citep{silich17}.  These models contrast
strongly with the standard paradigm for mechanical feedback driven by
energy-conserving superwinds.   Also, since any mechanical feedback 
takes time to mature, radiative feedback is expected to dominate at
the youngest SSC ages \citep[\eg][]{freyer03,krumholz09b}.  
Superwind suppression has been reported in SSCs in M82 \citep{smith06, westmoquette14} and NGC 2366 \citep{oey17}.

With the recent confirmation of high LyC escape fractions
(\fesclyc$=6-13$\%) from five out of five targeted ``Green Pea" (GP)
galaxies \citep{izotov16a,izotov16b}, we can now investigate these feedback models in the context of strong LyC emitters.
 As a population, the GPs show elevated
\oiii~$\lambda$5007/\oii~$\lambda$3727 ratios, as expected for
density-bounded nebulae \citep[\eg][]{jaskot13}. In addition to the confirmed leakers, several
other GPs show indirect evidence of LyC escape, such as weak
low-ionization absorption lines and narrow double-peaked
\lya\ profiles \citep{jaskot14,verhamme15,henry15}. Observations of
the five confirmed leakers show that \fesclyc\ may correlate with
\oiii/\oii, \sigsfr, and \lya\ properties such as
escape fraction (\fesclya), \lya\ equivalent width (EW), and separation of
the \lya\ peaks (\vpeaks; \citealt{izotov16b, verhamme17}).  

The role of mechanical feedback
in enabling LyC escape from the GPs is not yet clear. Some GPs show broad emission line wings
\citep[\eg][]{amorin12b}, which may suggest
that concentrated SN feedback is driving outflows. On the other hand, such broad
wings are not obviously due to
expanding shells or winds and are difficult to interpret
\citep{binette09}.  Furthermore,
\citet{henry15} find no correlation between low-ionization gas outflow
velocities and \fesclya, and \citet{chisholm17} show that LyC leaker outflow velocities are no more extreme than a control sample. 

Here, we present new, high S/N {\it Hubble Space Telescope (HST)}
Cosmic Origins Spectrograph (COS) observations of 13
extreme GPs, with \oiii/\oii$\sim7-34$. Combined with previous GP
samples \citep{jaskot14, henry15, chisholm17}, we assess whether gas velocities correlate with indirect diagnostics of
LyC escape and examine evidence for suppressed superwinds. In contrast to the prevailing picture of mechanical
feedback, we show that the best LyC emitter candidates have nearly
static gas velocities. 

\section{Sample and Measurements}
\label{sec:measure}
We selected a sample of 13 star-forming GPs at $z=0.027-0.123$ based on
high \oiii/\oii\ and high S/N from the Sloan Digital Sky Survey (SDSS) Data Release 10 (DR10; \citealt{ahn14}). We
correct all optical nebular lines for Milky Way extinction using the
\citet{schlafly11} maps and \citet{fitzpatrick99} attenuation
law. Following \citet{izotov17a}, we correct for internal dust
extinction using the \citet{cardelli89} law. We obtain COS spectra
with the G130M grating; we will describe these
observations in a forthcoming work (Jaskot \etal\ in prep.), but we summarize the main
points below. We bin the spectra to $12-34$ \kmps\ resolutions as
estimated from the full width at half maximum of the cross-dispersion
profiles. The spatial profiles are compact; we
estimate that deviations from a point-source profile will cause at
most a 6\% flux uncertainty per pixel in the final spectrum, and we incorporate this
uncertainty accordingly. The \lya\ spatial profiles are typically only
a few pixels more extended than the continuum emission. However,
because of our sources' low redshifts ($z\leq0.12$), some
\lya\ emission may be scattered outside the COS aperture
and missed entirely. 

To measure gas velocities, we first linearly fit the continuum region
near each spectral line and normalize the spectrum. Following \citet{henry15}, we measure a
characteristic velocity (\vchar), the velocity weighted by absorption
depth; and a maximum velocity (\vmax), the velocity where the
absorption meets the continuum. For GPs from
\citet{jaskot14} and the new sample, we measure velocities for the
low-ionization lines \oi~$\lambda$1302, \sitwo~$\lambda\lambda$1190,
1193, 1260, 1304, and \cii~$\lambda$1334; and the high-ionization lines
\sithree~$\lambda$1206 and \siiv~$\lambda\lambda$1394,1403. Not all
lines are covered or detected in each object; some spectra only cover the low-ionization lines. The highest \oiii/\oii\ GP
in our sample, J160810+352809 (J1608), shows the high ionization lines in emission with
superimposed absorption forming double-peaked line profiles. For this
object, we set \vchar\ to the velocity of the deepest absorption and
\vmax\ equal to the velocity of the blue emission peak. For the LyC-emitting GPs, we adopt the $v_{\rm cen}$ measurements from \citet{chisholm17}, which are comparable to \vchar. 

\begin{table*}
\centering
\caption{GP Sample Properties}
\begin{tabular}{lccccccccc}
\hline
\hline
& & & & & \multicolumn{3}{c}{Low-ionization\tablenotemark{a}} & \multicolumn{2}{c}{High-ionization\tablenotemark{a}} \\
Galaxy & $z$ & $\frac{\rm [O\thinspace{\sc III}]}{\rm [O\thinspace{\sc II}]}$ & \fesclya & \vpeaks & EW\tablenotemark{b} &  \vchar & \vmax & \vchar & \vmax \\
 & & &  & (\kmps) & (\AA) & (\kmps) & (\kmps) & (\kmps) & (\kmps) \\
\hline
J144805-011058 & 0.0274 & 7.3 &  0.00 & --- & -1.04 (6) & -5 (6) & -281 (5) & -136 (3) & -435 (3) \\
J150934+373146 & 0.0326 & 13.8 &  0.04$\pm$0.02 & 400$\pm$27 & -0.35 (5) & -4 (5) & -118 (5) & -67 (3) & -198 (3) \\
J160810+352809 & 0.0327 & 33.0 &  0.16$\pm$0.04 & 214$\pm$30 & -0.17 (3) & 19 (3) & -95 (3) & 2 (3) & -46 (3)  \\
J230210+004939 & 0.0331 & 8.3 &  0.26$\pm$0.06 & 279$\pm$48 & -0.15 (3) & -28 (3) & -116 (3) & -50 (3) & -230 (3)  \\
J021307+005612 & 0.0399 & 6.9 &  0.11$\pm$0.02 & 397$\pm$47 & -0.48 (6) & -29 (6) & -203 (6) & -82 (1) & -315 (1) \\
J080841+172856 & 0.0442 & 9.7  & 0.31$\pm$0.07 & 156$\pm$37; 441$\pm$58 & --- & --- & --- & -236 (1) & -324 (1)  \\
J173501+570309 & 0.0472 & 6.8 & 0.09$\pm$0.02 & 460$\pm$47 & -0.53 (6) & -85 (6) & -401 (5) & -156 (1) & --- \\
J131131-003844 & 0.0811 & 6.7 &  0.24$\pm$0.05 & 273$\pm$26 & -0.32 (6) & -83 (6) & -233 (5) & --- & ---  \\
J120016+271959 & 0.0819 & 9.0 &  0.40$\pm$0.08 & 327$\pm$65 & -0.17 (3) & -89 (3) & -214 (3) & --- & --- \\
J024052-082827 & 0.0822 & 14.1 &  0.20$\pm$0.04 & 266$\pm$29 & -0.58 (6) & -149 (6) & -412 (5) & --- & ---  \\
J085116+584055 & 0.0919 & 9.1 &  0.04$\pm$0.01 & 361$\pm$25 & -0.64 (5) & -44 (5) & -186 (5) & --- & ---  \\
J122612+041536 & 0.0942 & 7.7 &  0.11$\pm$0.02 & 360$\pm$40 & -0.41 (4) & -25 (4) & -137 (3) & --- & ---  \\
J133538+080149 & 0.1235 & 6.8 &  0.00 & --- & -1.04 (5) & 3 (5) & -170 (5) & --- & ---  \\
J081552+215624 & 0.1410 & 9.9  & 0.27$\pm$0.06 & 296$\pm$52 & --- & --- & --- & -318 (1) & ---  \\
J145735+223202 & 0.1487 & 7.2  & 0.01$\pm$0.02 & 749$\pm$56& -0.79 (4) & -39 (4) & -299 (3) & -155 (3) & -189 (1)  \\
J030321-075923 & 0.1649 & 6.8  & 0.05$\pm$0.01 & 443$\pm$156 & -0.50\tablenotemark{c} (4) & -200\tablenotemark{c} (2) & -560\tablenotemark{c} (1) & -240\tablenotemark{c} (3) & -580\tablenotemark{c} (3)  \\
J092600+442737 & 0.1807 & 3.1  & 0.20$\pm$0.03 & 409$\pm$71\tablenotemark{c} & -0.45\tablenotemark{c} (4) & -280\tablenotemark{c} (2) & -580\tablenotemark{c} (2) & -320\tablenotemark{c} (3) & -880\tablenotemark{c} (3)  \\
J142406+421646 & 0.1848 & 5.7  & 0.24$\pm$0.04 & 374$\pm$71\tablenotemark{c} & -0.50\tablenotemark{c} (2) & -235\tablenotemark{c} (2) & -480\tablenotemark{c} (2) & -280\tablenotemark{c} (1) & -560\tablenotemark{c} (1)  \\
J113722+352427 & 0.1944 & 2.7  & 0.11$\pm$0.02 & 556$\pm$71\tablenotemark{c}& -1.15\tablenotemark{c} (4) & -145\tablenotemark{c} (4) & -445\tablenotemark{c} (4) & -170\tablenotemark{c} (3) & -510\tablenotemark{c} (3)  \\
J121904+152609 & 0.1956 & 10.3  & 0.55$\pm$0.08 & 242$\pm$43 & ---  & --- & --- & -395\tablenotemark{c} (2) & -815\tablenotemark{c} (2) \\
J124423+021540 & 0.2394 & 3.7  & 0.07$\pm$0.01 & 487$\pm$71\tablenotemark{c} & -1.20\tablenotemark{c} (4) & -90\tablenotemark{c} (4) & -380\tablenotemark{c} (4) & -100\tablenotemark{c} (3) & -400\tablenotemark{c} (3)  \\
J113304+651341 & 0.2414 & 3.5  & 0.27$\pm$0.05 & 340$\pm$71\tablenotemark{c} & --- & ---  & --- & -270\tablenotemark{c} (1) & ---   \\
J105331+523753 & 0.2526 & 2.3  & 0.06$\pm$0.01 & 506$\pm$71\tablenotemark{c} & -0.80\tablenotemark{c} (3) & -140\tablenotemark{c} (3) & -440\tablenotemark{c} (3) & -180\tablenotemark{c} (3) & -650\tablenotemark{c} (3)  \\
J091113+183108 & 0.2622 & 1.8  & 0.16$\pm$0.03 & 359$\pm$71\tablenotemark{c} & -0.80\tablenotemark{c} (4) & -250\tablenotemark{c} (4) & -525\tablenotemark{c} (4) & -290\tablenotemark{c} (3) & -815\tablenotemark{c} (2)  \\
J124835+123403 & 0.2634 & 3.4  & 0.42$\pm$0.06 & --- & -0.85\tablenotemark{c} (2) & -180\tablenotemark{c} (2) & -420\tablenotemark{c} (3) & -250\tablenotemark{c} (3) & -470\tablenotemark{c} (3)  \\
J144231-020952 & 0.2937 & 6.7 & 0.51$\pm$0.03 & 310$\pm$14\tablenotemark{d} & -0.65\tablenotemark{e} (1) & -473\tablenotemark{e} (1) & --- & -250\tablenotemark{e} (1) & --- \\
J092532+140313 & 0.3013 & 4.8 & 0.30$\pm$0.01 & 310$\pm$14\tablenotemark{d} &  -0.75\tablenotemark{e} (1) & -64\tablenotemark{e} (1) & --- & -135\tablenotemark{e} (1) & --- \\
J133304+624604 & 0.3181 & 4.8 & 0.54$\pm$0.04 & 390$\pm$14\tablenotemark{d} & -0.75\tablenotemark{e} (1) & -237\tablenotemark{e} (1) & --- & -311\tablenotemark{e} (1) & --- \\
J115205+340050 & 0.3419 & 5.4 & 0.34$\pm$0.02 & 270$\pm$14\tablenotemark{d} & -0.49\tablenotemark{e} (1) & -336\tablenotemark{e} (1) & --- & -193\tablenotemark{e} (1) & --- \\
J150343+364451 & 0.3557 & 4.9 & 0.29$\pm$0.02 & 430$\pm$14\tablenotemark{d} & -0.56\tablenotemark{e} (1) & -187\tablenotemark{e} (1) & --- & -128\tablenotemark{e} (1) & --- \\
\hline
\end{tabular}
\tablenotetext{$a$}{Median values with the number of observed lines in parentheses. Typical uncertainties on individual measurements are 0.19 \AA, 41 \kmps, and 32 \kmps\ for low-ionization EW, \vchar, and \vmax\ and 54 \kmps\ and 51 \kmps\ for high-ionization \vchar\ and \vmax. } \tablenotetext{$b$}{Negative EWs indicate absorption.} \tablenotetext{$c$}{From \citep{henry15}.} \tablenotetext{$d$}{From \citep{verhamme17}.} \tablenotetext{$e$}{From \citep{chisholm17}.}
\label{tab:sample}
\end{table*}

Each velocity
measurement is subject to different systematic effects. Low spectral resolution or 
infilling by scattered emission may shift \vchar\ to more negative
velocities, while low S/N or low EW absorption may lead to
underestimates of \vmax. The
\siiv~$\lambda\lambda1393,1403$ doublet could also contain significant stellar absorption. However, the measured \siiv\ velocities do not differ systematically from the velocities of the interstellar \sithree\ absorption, with median differences of only 30 \kmps\ in \vchar\ and 110 \kmps\ in \vmax.

We compare the observed velocities with three indirect measures of \hi\ optical depth:
\vpeaks, the median EW of the low-ionization absorption
lines listed above, and \fesclya\ (Table~\ref{tab:sample}). The velocity separation of the red and blue peaks of a double-peaked
\lya\ spectral profile should correlate with optical depth, as
reduced \lya\ scattering at low \nhi\ produces a narrower spectral
profile \citep{verhamme15, dijkstra16}. For the galaxy J080841+172856 (J0808), which has a triple-peaked profile, we calculate
\vpeaks\ twice, once for each blue peak. We also calculate the EW
of low-ionization absorption lines, which should be weaker in LyC
emitters \citep[\eg][]{heckman01}, although gas metallicity and
emission scattered into the line of sight will also affect the
observed EW. 

\begin{sloppypar}Galaxies with low \nhi\ should also have high
\fesclya\ \citep[\eg][]{yajima14,dijkstra16,verhamme17}. However, aperture effects may lower
\fesclya\ in the new $z<0.12$ GPs compared to other GP samples
\citep{jaskot14,henry15,izotov16b}. To obtain \fesclya, we measure the
\lya\ emission within any associated \lya\ absorption trough; the uncertainties account for the difference between including and excluding absorption. After correcting the 
flux for Milky Way extinction, we derive the intrinsic \lya\ flux
before scattering by scaling the extinction-corrected \halpha\ flux by
the Case B \lya/\halpha\ ratio \citep{dopita03} at the starburst's
calculated electron temperature and density (\lya/\halpha$=8.24-8.96$ for our sample). Deviations from Case B only become important at \fesclyc $>$ 90\% \citep{ferland99}. For consistency, we
re-calculate \fesclya\ for the \citet{henry15} and \citet{izotov16b}
GPs in the same manner. \end{sloppypar}

\section{Results}
\label{sec:results}
\subsection{Velocity and Optical Depth}
\label{sec:velocities}
\begin{figure*}
\epsscale{1.2}
\plotone{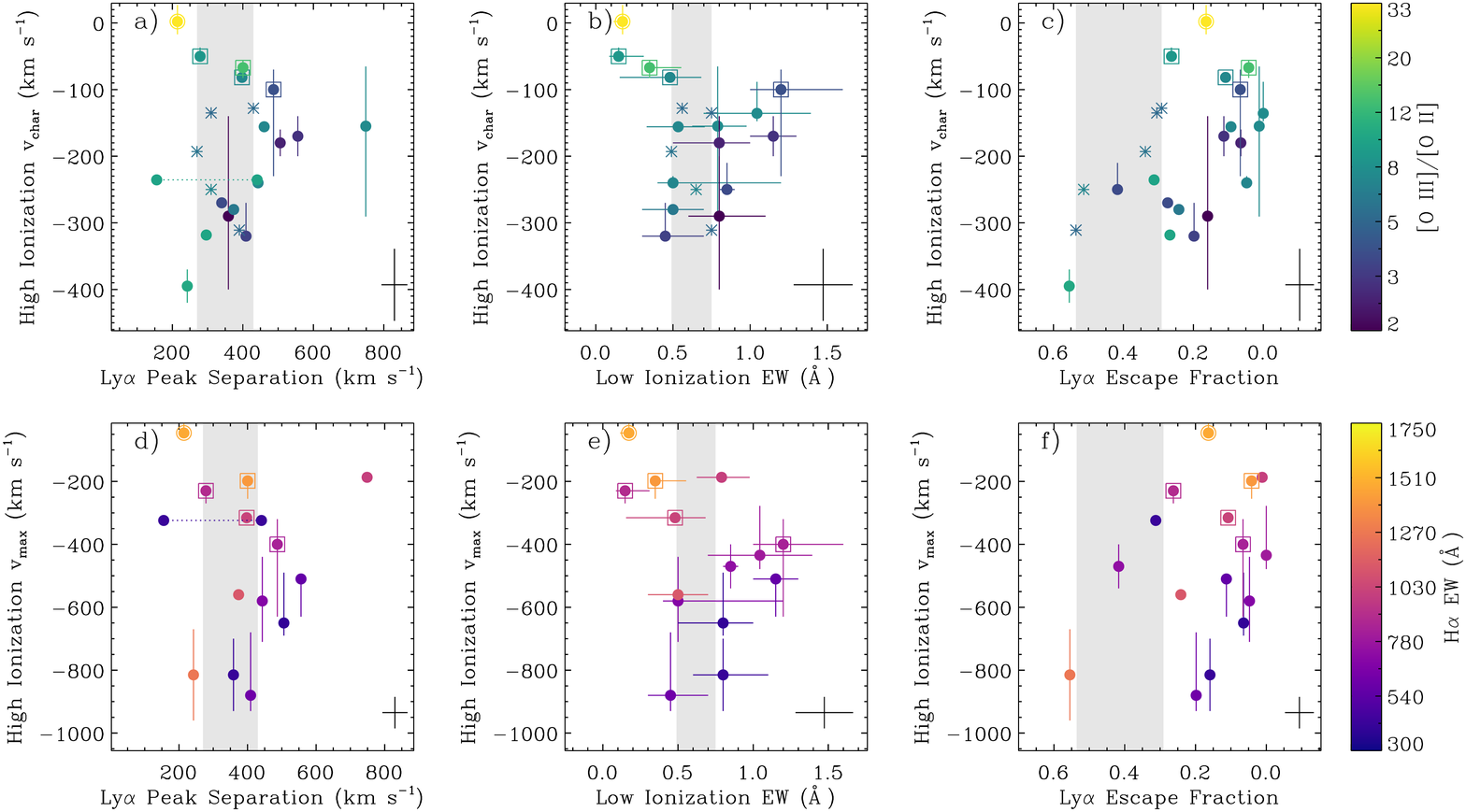}
\caption{Top panels show the median \vchar\ of high-ionization absorption lines (\sithree~$\lambda$1206 and \siiv~$\lambda\lambda$1394,1403) as a function of indirect optical depth diagnostics (a: \vpeaks, b: median low-ionization EW, and c: \fesclya). Color indicates \oiii/\oii. Bottom panels show the median \vmax\ of the high-ionization lines as a function of the same diagnostics, with color indicating H$\alpha$ EW. Optical depth increases to the right in all panels. Data points and error bars show the median and range, respectively, of the lines observed in a given GP. The objects with the five lowest \vchar\ values are surrounded by open symbols and suggest an inverse trend, with  slower velocities at lower optical depths; the open circle indicates J1608. For J0808's triple-peaked \lya\ profile, we show both values of \vpeaks\ connected with a dotted line. The black error bar in the corner indicates the typical errors on individual line measurements. Gray bands show the range of \vpeaks, \sitwo~$\lambda$1260 EWs, and \fesclya\ observed in the confirmed LyC-emitting GPs \citep{verhamme17, chisholm17}, which are plotted with asterisks. }
\label{fig:highi}
\end{figure*}

In Figure~\ref{fig:highi}, we show optical depth
diagnostics as a function of the measured velocities of
high-ionization gas, characteristic of a starburst's wind or wind interface regions. For clarity, we plot one data point for each GP, showing the median and range of the observed absorption lines. In Figure~\ref{fig:highi}, most of the objects follow a
correlation between blue-shifted high-ionization gas velocity and
diagnostics of lower LyC optical depth.  However, the subset of objects
with the lowest \vchar\ appears to show the {\it opposite} trend:
at lower optical depths, the gas velocities
decrease. 

These objects include the highest excitation GPs. The best LyC-leaker candidates, with low \vpeaks\ and weak low-ionization EWs, generally have high \oiii/\oii\ ratios and high H$\alpha$ EWs \citep[\cf][]{izotov16b}. These same galaxies tend to have low outflow velocities. The GPs with 
outflow velocities $<100$ \kmps\ all have \vpeaks$<400$ \kmps, as narrow as known GP LyC emitters
\citep{verhamme17}. With burst ages of only a few Myr \citep[\eg][]{jaskot13,izotov17b}, some GPs may be too young for SN feedback, although stellar winds should still be present. These results support predictions that superwinds may be inhibited by catastrophic cooling and high pressure in extreme SSCs \citep[\eg][]{silich07, silich17}. These extreme SSCs may have conditions that promote LyC escape, such as high ionizing fluxes, combined with conditions detrimental to superwind formation. Radiative, not mechanical, feedback may dominate in young, compact starbursts like the GPs \citep[\eg][]{freyer03,krumholz09b}. The remaining, typically lower excitation, objects do suggest a correlation between velocity and low line-of-sight optical depth. These starbursts may not have suppressed superwinds or may be dominated by SN activity. 

\begin{figure*}
\epsscale{1.2}
\plotone{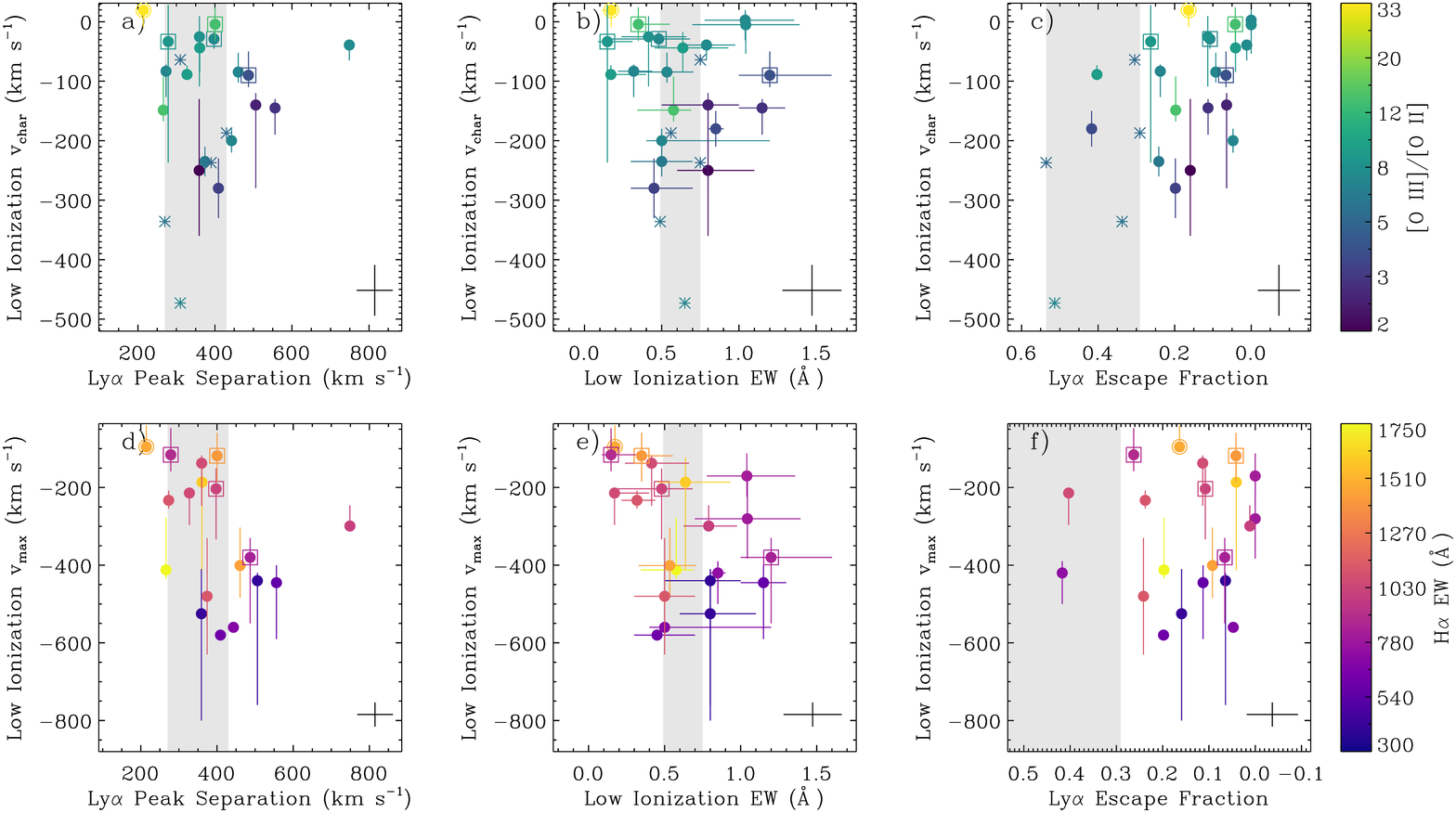}
\caption{The panels and symbols are the same as Figure~\ref{fig:highi} but for the velocities of low-ionization absorption lines (\oi~$\lambda$1302, \sitwo~$\lambda\lambda$1190, 1193, 1260, 1304, and \cii~$\lambda$1334).}
\label{fig:lowi}
\end{figure*}

The low-ionization gas, which tracks neutral gas kinematics, could originate near or far from the starburst and may not necessarily trace the starburst's wind. However, this gas will have the greatest effect on \lya\ and LyC escape. Many objects again appear to show both low velocities and indirect indicators of low optical depth (Figure~\ref{fig:lowi}).  As with the high-ionization gas, the GPs
with narrow \lya\ tend to have the slowest outflow velocities (Figure~\ref{fig:lowi}a,d), and most GPs with weak low-ionization
absorption show low velocities in both \vchar\ and \vmax.  

Systematic effects are unlikely to alter these conclusions. Programs GO-12928 \citep{henry15} and GO-13293 \citep{jaskot14} include GPs with a range of \oiii/\oii\ ratios from the original $z>0.1$ GP sample of \citet{cardamone09}, while we selected our new sample (GO-14080) based on observability and high \oiii/\oii. These new GPs have systematically higher \oiii/\oii\ ratios and lower redshifts than the previous samples. Since the optical depth in \lya\ is $10^4$ times than that of LyC, \lya\ will scatter even in LyC emitters. The COS aperture will capture less of this scattered \lya\ halo in lower redshift GPs. At $z=0.047$, the median redshift of the new sample, the COS aperture subtends 2.3 kpc, $\sim1/5$ the aperture size for the \citep{izotov16b} LyC leakers, and at a given \vpeaks, GPs with $z<0.1$ show systematically lower \fesclya. The \fesclya\ values for these GPs are likely underestimated; higher \fesclya\ values would only increase the number of GPs with low optical depth and low velocities in Figures~\ref{fig:highi}c,f and \ref{fig:lowi}c,f. 

For the higher-redshift GPs, metal line emission captured by the larger physical aperture could fill in absorption at the systemic velocity, thereby causing \vchar\ to appear more blue-shifted. However, we observe the same trends with \vmax, which should be less affected by scattered emission. Even if emission is biasing \vchar\ too high, this effect merely emphasizes that the GPs' velocities must truly be low. Lastly, we could underestimate \vmax\ due to low S/N. However, the high \oiii/\oii\ GPs show the lowest \vmax\ values yet generally have higher S/N spectra. At low optical depths, weak low-ionization absorption could lead to underestimated \vmax\ values, but the low \vmax\ values appear in the stronger high-ionization absorption lines as well. We show a characteristic example of some of the weakest detected low-ionization absorption lines in Figure~\ref{fig:j1608_threepanel}. Although the lines are weak, they are narrow and centered at systemic velocity; a significantly higher \vmax\ seems unlikely.

Although outflows are not driving the GPs' low optical depths, they may have a slight effect on \lya\ escape. In Figures~\ref{fig:highi}c,f and \ref{fig:lowi}c,f, only \fesclya\ may show an overall trend with faster outflows. This tentative trend may arise because \lya\ photons are out of resonance with the higher velocity gas and thereby escape more easily \citep{mashesse03}. However, the correlation is most evident in the high-ionization gas velocities, not the low-ionization gas that should affect \lya\ escape \citep[\cf][]{henry15}. As discussed above, aperture effects may account for part of the observed \fesclya\ trend by leading to systematic underestimates of \fesclya\ for the highest \oiii/\oii\ GPs. For low-redshift galaxies, \fesclya\ may be an unreliable indicator of LyC escape. 

\subsection{J1608: Optical Depth at Extreme Ionization}
\label{sec:j1608}
\begin{figure*}
\epsscale{1.2}
\plotone{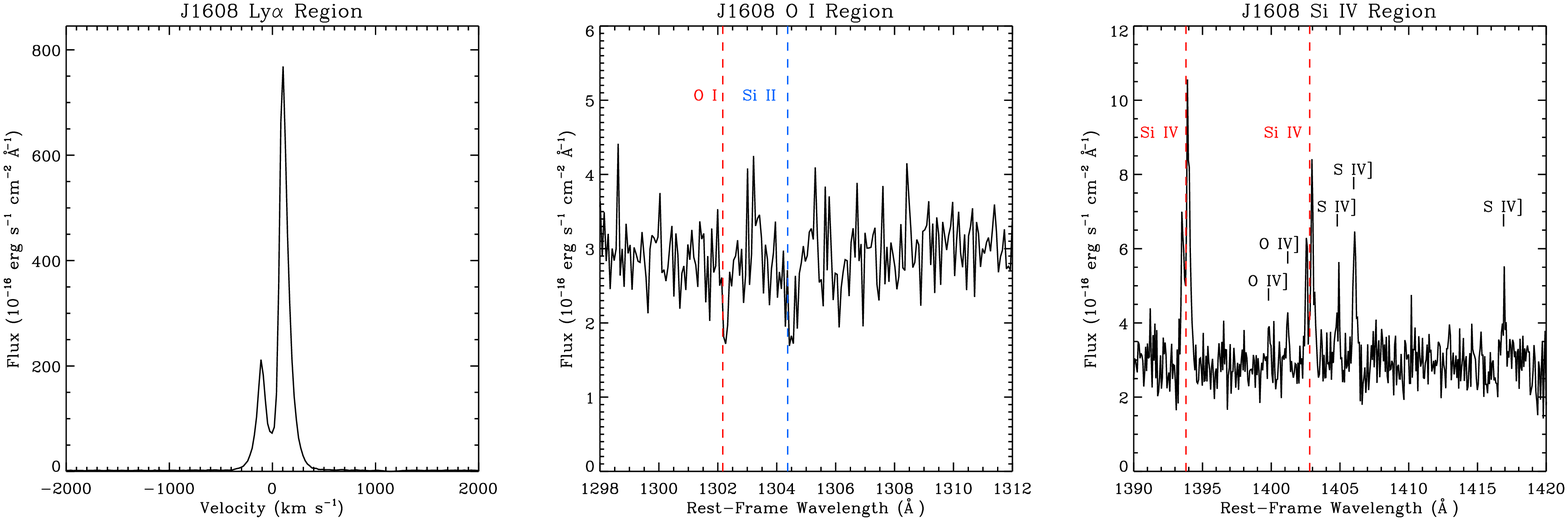}
\caption{J1608's \lya\ profile is strong and narrow with significant flux at the systemic velocity ({\it left panel}). Its low-ionization absorption lines ({\it central panel}) are weak, narrow, and barely offset from the expected transition wavelength (dashed lines). J1608 shows high ionization lines such as \siiv\ ({\it right panel}) and \sithree\ in emission with absorption superimposed at the systemic velocity. Dashed red lines indicate the \siiv\ transition wavelengths; labels identify other lines from \oivp\ and \sivp.}
\label{fig:j1608_threepanel}
\end{figure*}

\begin{figure}
\plotone{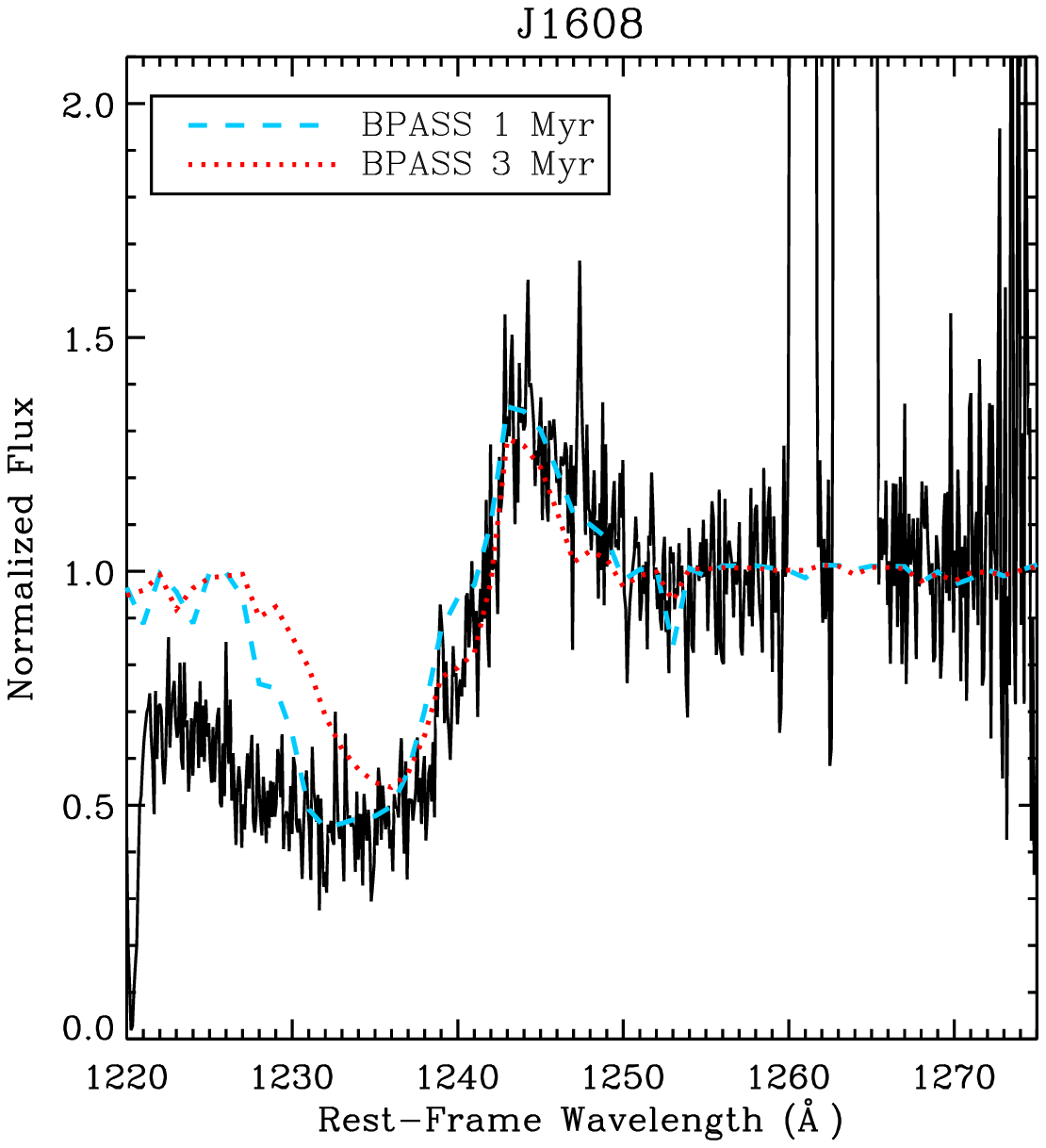}
\caption{We compare the \nv~$\lambda$1240 line in J1608 with predictions from the BPASS models \citep{stanway16} for a 1 Myr and 3 Myr instantaneous burst with a 150\Msol\ upper mass limit. The height and depth of the \nv\ line suggest an age close to 1 Myr. \lya\ absorption affects the blue side of the observed \nv\ profile.}
\label{fig:j1608_nv}
\end{figure}

\begin{sloppypar} The most extreme galaxy in the sample, J1608, has the highest \oiii/\oii\ ratio among SDSS DR10 star-forming galaxies and exemplifies the disconnect between gas velocity and \nhi. With a dust-corrected \oiii/\oii$=33$, J1608's enormous ionization parameter may indicate a density-bounded nebula \citep[\eg][]{jaskot13}, a conclusion supported by other spectral diagnostics \citep{izotov17b}. J1608 is likely extremely young; its rest-frame H$\alpha$ EW is 1410 \AA, and its strong \nv~$1240$\AA\ stellar P-Cygni feature matches a $\sim$1 Myr stellar population (Figure~\ref{fig:j1608_nv}) with Very Massive Stars ($M>100$\Msol). Furthermore, J1608 shows the unprecedented detection of \sithree~$\lambda1206$ and \siiv~$\lambda\lambda$1394,1403 as double-peaked emission lines, along with fainter emission from semi-forbidden \oivp\ and \sivp\ (Figure~\ref{fig:j1608_threepanel}). CLOUDY photoionization models \citep{ferland98,jaskot16} predict strong nebular \siiv\ emission only for  $\log(U) > -2$, while stellar \siiv\ absorption weakens with both age and metallicity \citep[\eg][]{stanway16}. The fact that \siiv\ emission dominates absorption suggests that J1608 must be young and highly ionized and is consistent with its low metallicity ($12+\log(\rm{O/H})=7.79$; \citealt{izotov17b}). 
\end{sloppypar}

J1608 shows no evidence of SN-driven outflows, yet it is also the best LyC emitter candidate in our
sample. Its absorption lines are nearly static, with \vchar\ ranging
from $-17$ to $+27$
\kmps\ (Figure~\ref{fig:j1608_threepanel}). Nevertheless, J1608 may have one of the lowest column densities for
low-redshift starbursts. Its low-ionization absorption EWs are all
$<0.2$\AA, and its \lya\ peaks are separated by 214
\kmps\ (Figure~\ref{fig:j1608_threepanel}), lower than any other
double-peaked \lya\ profile yet reported. J1608's \lya\ EW ($\sim$163 \AA) is among the strongest of the
GPs, and J1608 has more \lya\ flux
at the systemic velocity, where \lya\ optical depth should be
highest. Although J1608's \fesclya\ of 0.16 is intermediate for the
sample, the observations may experience more aperture losses given J1608's low redshift ($z\sim0.03$). Dust may also suppress scattered \lya\ photons,
as J1608's $E(B-V)=0.20$ is above the sample average. The evidence from 
both the \lya\ profile and the low-ionization lines strongly suggests
low \nhi\ along the line of sight. J1608 offers a clear template of
the extreme GP properties linked to LyC emission.

\section{Discussion}
\label{sec:discussion}
Our results support predictions for an extreme feedback
regime characterized by suppressed superwinds. In extreme, young starbursts like the GPs, catastrophic cooling and high pressure may prevent stellar winds and supernovae from forming a superwind \citep[\eg][]{silich07, silich17}.
In addition, we may be catching these
starbursts at particularly young ages, $< 3$ Myr, before the death of
the most massive stars as SNe. The GPs' optical spectra show enormous
Balmer line EWs \citep{jaskot13}, unusually strong nebular continua
\citep{izotov11}, and nebular emission from the weak, age-sensitive
\hei~$\lambda$3819 and higher-order Balmer series lines
\citep{jaskot13}. The strong stellar \nv~$\lambda$1240 P-Cygni profile
observed in the extreme GP J1608 further indicates an age $<3$ Myr
(\S~\ref{sec:j1608}). With young ages, the GPs may be at
their peak LyC production \citep[\cf][]{schaerer16,izotov17a}. By
ionizing the surrounding nebular gas, this intense radiation may
ensure its own escape. 

\begin{sloppypar}The recently identified, local GP analog Mrk 71-A \citep{micheva17} appears to be a similarly young starburst without evident
mechanical feedback. Mrk 71-A may be only $\sim$1
Myr-old, and CO kinematics demonstrate the absence of an energy-driven superwind from the $10^5$ \Msol\ SSC
\citep{oey17}. Radiation may
dominate Mrk 71-A's feedback, and a high star-formation efficiency
may also contribute to the removal of neutral gas \citep{oey17}. Like the GPs, Mrk 71-A's optical emission lines show
mysterious broad wings, but SN activity may not generate
this feature \citep[\eg][]{roy92, binette09}. Instead, these wings could be associated with the catastrophic cooling of the mechanical feedback \citep[\eg][]{tenorio10}.
 \end{sloppypar}

Our results suggest that LyC emitters may come in different types, with
suppressed superwinds in the most extreme GP systems,
and conventional wind and SN feedback clearing out neutral gas in other
systems. One known high-redshift LyC
emitter is GP-like \citep{vanzella16}, while at least one other is not
\citep{shapley16}. Ultimately, the reionization of the universe may
require the contribution of different types of galaxies
\citep[\eg][]{fontanot12}. Extremely young starbursts, like the GPs,
may contribute thanks to prolific LyC production combined with strong
radiative feedback. As these starbursts age, SN feedback may provide a
new means of LyC escape, while older sources, such as evolved binary
systems, contribute the LyC photons \citep[\eg][]{ma16,stanway16}. The
existence of multiple mechanisms for LyC escape may help explain how
galaxies were able to reionize the universe. 

\acknowledgments
We thank the referee for a helpful response and thank Crystal Martin for useful discussions. AEJ acknowledges support by NASA through Hubble Fellowship grant HST-HF2-51392 and STScI grant HST-GO-14080. MSO acknowledges support from NASA through grant HST-GO-14080 from STScI. STScI is operated by AURA under NASA contract NAS-5-26555. TD acknowledges support from the Massachusetts Space Grant Consortium. Funding for SDSS-III has been provided by the Alfred P. Sloan Foundation, the Participating Institutions, the National Science Foundation, and the U.S. Department of Energy Office of Science. SDSS-III is managed by the Astrophysical Research Consortium for the Participating Institutions of the SDSS-III Collaboration.



\begin{thebibliography}{}
\expandafter\ifx\csname natexlab\endcsname\relax\def\natexlab#1{#1}\fi
\providecommand{\url}[1]{\href{#1}{#1}}

\bibitem[{{Ahn} {et~al.}(2014){Ahn}, {Alexandroff}, {Allende Prieto}, {Anders},
  {Anderson}, {Anderton}, {Andrews}, {Aubourg}, {Bailey}, {Bastien}, \&
  et~al.}]{ahn14}
{Ahn}, C.~P., {Alexandroff}, R., {Allende Prieto}, C., {et~al.} 2014, \apjs,
  211, 17

\bibitem[{{Alexandroff} {et~al.}(2015){Alexandroff}, {Heckman}, {Borthakur},
  {Overzier}, \& {Leitherer}}]{alexandroff15}
{Alexandroff}, R.~M., {Heckman}, T.~M., {Borthakur}, S., {Overzier}, R., \&
  {Leitherer}, C. 2015, \apj, 810, 104

\bibitem[{{Amor{\'{\i}}n} {et~al.}(2012){Amor{\'{\i}}n}, {V{\'{\i}}lchez},
  {H{\"a}gele}, {Firpo}, {P{\'e}rez-Montero}, \& {Papaderos}}]{amorin12b}
{Amor{\'{\i}}n}, R., {V{\'{\i}}lchez}, J.~M., {H{\"a}gele}, G.~F., {et~al.}
  2012, \apjl, 754, L22

\bibitem[{{Binette} {et~al.}(2009){Binette}, {Drissen}, {Ubeda}, {Raga},
  {Robert}, \& {Krongold}}]{binette09}
{Binette}, L., {Drissen}, L., {Ubeda}, L., {et~al.} 2009, \aap, 500, 817

\bibitem[{{Cardamone} {et~al.}(2009){Cardamone}, {Schawinski}, {Sarzi},
  {Bamford}, {Bennert}, {Urry}, {Lintott}, {Keel}, {Parejko}, {Nichol},
  {Thomas}, {Andreescu}, {Murray}, {Raddick}, {Slosar}, {Szalay}, \&
  {Vandenberg}}]{cardamone09}
{Cardamone}, C., {Schawinski}, K., {Sarzi}, M., {et~al.} 2009, \mnras, 399,
  1191

\bibitem[{{Cardelli} {et~al.}(1989){Cardelli}, {Clayton}, \&
  {Mathis}}]{cardelli89}
{Cardelli}, J.~A., {Clayton}, G.~C., \& {Mathis}, J.~S. 1989, \apj, 345, 245

\bibitem[{{Chisholm} {et~al.}(2017){Chisholm}, {Orlitov{\'a}}, {Schaerer},
  {Verhamme}, {Worseck}, {Izotov}, {Thuan}, \& {Guseva}}]{chisholm17}
{Chisholm}, J., {Orlitov{\'a}}, I., {Schaerer}, D., {et~al.} 2017, \aap, 605,
  A67

\bibitem[{{Clarke} \& {Oey}(2002)}]{clarke02}
{Clarke}, C., \& {Oey}, M.~S. 2002, \mnras, 337, 1299

\bibitem[{{Dijkstra} {et~al.}(2016){Dijkstra}, {Gronke}, \&
  {Venkatesan}}]{dijkstra16}
{Dijkstra}, M., {Gronke}, M., \& {Venkatesan}, A. 2016, \apj, 828, 71

\bibitem[{{Dopita} \& {Sutherland}(2003)}]{dopita03}
{Dopita}, M.~A., \& {Sutherland}, R.~S. 2003, {Astrophysics of the diffuse
  universe}

\bibitem[{{Ferland}(1999)}]{ferland99}
{Ferland}, G.~J. 1999, \pasp, 111, 1524

\bibitem[{{Ferland} {et~al.}(1998){Ferland}, {Korista}, {Verner}, {Ferguson},
  {Kingdon}, \& {Verner}}]{ferland98}
{Ferland}, G.~J., {Korista}, K.~T., {Verner}, D.~A., {et~al.} 1998, \pasp, 110,
  761

\bibitem[{{Fitzpatrick}(1999)}]{fitzpatrick99}
{Fitzpatrick}, E.~L. 1999, \pasp, 111, 63

\bibitem[{{Fontanot} {et~al.}(2012){Fontanot}, {Cristiani}, \&
  {Vanzella}}]{fontanot12}
{Fontanot}, F., {Cristiani}, S., \& {Vanzella}, E. 2012, \mnras, 425, 1413

\bibitem[{{Freyer} {et~al.}(2003){Freyer}, {Hensler}, \& {Yorke}}]{freyer03}
{Freyer}, T., {Hensler}, G., \& {Yorke}, H.~W. 2003, \apj, 594, 888

\bibitem[{{Heckman} {et~al.}(2001){Heckman}, {Sembach}, {Meurer}, {Leitherer},
  {Calzetti}, \& {Martin}}]{heckman01}
{Heckman}, T.~M., {Sembach}, K.~R., {Meurer}, G.~R., {et~al.} 2001, \apj, 558,
  56

\bibitem[{{Henry} {et~al.}(2015){Henry}, {Scarlata}, {Martin}, \&
  {Erb}}]{henry15}
{Henry}, A., {Scarlata}, C., {Martin}, C.~L., \& {Erb}, D. 2015, \apj, 809, 19

\bibitem[{{Izotov} {et~al.}(2017{\natexlab{a}}){Izotov}, {Guseva}, {Fricke},
  {Henkel}, \& {Schaerer}}]{izotov17a}
{Izotov}, Y.~I., {Guseva}, N.~G., {Fricke}, K.~J., {Henkel}, C., \& {Schaerer},
  D. 2017{\natexlab{a}}, \mnras, 467, 4118

\bibitem[{{Izotov} {et~al.}(2011){Izotov}, {Guseva}, \& {Thuan}}]{izotov11}
{Izotov}, Y.~I., {Guseva}, N.~G., \& {Thuan}, T.~X. 2011, \apj, 728, 161

\bibitem[{{Izotov} {et~al.}(2016{\natexlab{a}}){Izotov}, {Orlitov{\'a}},
  {Schaerer}, {Thuan}, {Verhamme}, {Guseva}, \& {Worseck}}]{izotov16a}
{Izotov}, Y.~I., {Orlitov{\'a}}, I., {Schaerer}, D., {et~al.}
  2016{\natexlab{a}}, \nat, 529, 178

\bibitem[{{Izotov} {et~al.}(2016{\natexlab{b}}){Izotov}, {Schaerer}, {Thuan},
  {Worseck}, {Guseva}, {Orlitova}, \& {Verhamme}}]{izotov16b}
{Izotov}, Y.~I., {Schaerer}, D., {Thuan}, T.~X., {et~al.} 2016{\natexlab{b}},
  ArXiv e-prints, arXiv:1605.05160

\bibitem[{{Izotov} {et~al.}(2017{\natexlab{b}}){Izotov}, {Thuan}, \&
  {Guseva}}]{izotov17b}
{Izotov}, Y.~I., {Thuan}, T.~X., \& {Guseva}, N.~G. 2017{\natexlab{b}}, \mnras,
  471, 548

\bibitem[{{Jaskot} \& {Oey}(2013)}]{jaskot13}
{Jaskot}, A.~E., \& {Oey}, M.~S. 2013, \apj, 766, 91

\bibitem[{{Jaskot} \& {Oey}(2014)}]{jaskot14}
---. 2014, \apjl, 791, L19

\bibitem[{{Jaskot} \& {Ravindranath}(2016)}]{jaskot16}
{Jaskot}, A.~E., \& {Ravindranath}, S. 2016, \apj, 833, 136

\bibitem[{{Krause} \& {Diehl}(2014)}]{krause14}
{Krause}, M.~G.~H., \& {Diehl}, R. 2014, \apjl, 794, L21

\bibitem[{{Krumholz} \& {Matzner}(2009)}]{krumholz09b}
{Krumholz}, M.~R., \& {Matzner}, C.~D. 2009, \apj, 703, 1352

\bibitem[{{Ma} {et~al.}(2016){Ma}, {Hopkins}, {Kasen}, {Quataert},
  {Faucher-Gigu{\`e}re}, {Kere{\v s}}, {Murray}, \& {Strom}}]{ma16}
{Ma}, X., {Hopkins}, P.~F., {Kasen}, D., {et~al.} 2016, \mnras,
  arXiv:1601.07559

\bibitem[{{Ma} {et~al.}(2015){Ma}, {Kasen}, {Hopkins}, {Faucher-Gigu{\`e}re},
  {Quataert}, {Kere{\v s}}, \& {Murray}}]{ma15}
{Ma}, X., {Kasen}, D., {Hopkins}, P.~F., {et~al.} 2015, \mnras, 453, 960

\bibitem[{{Mas-Hesse} {et~al.}(2003){Mas-Hesse}, {Kunth}, {Tenorio-Tagle},
  {Leitherer}, {Terlevich}, \& {Terlevich}}]{mashesse03}
{Mas-Hesse}, J.~M., {Kunth}, D., {Tenorio-Tagle}, G., {et~al.} 2003, \apj, 598,
  858

\bibitem[{{Micheva} {et~al.}(2017){Micheva}, {Oey}, {Jaskot}, \&
  {James}}]{micheva17}
{Micheva}, G., {Oey}, M.~S., {Jaskot}, A.~E., \& {James}, B.~L. 2017, \apj,
  845, 165

\bibitem[{{Oey} {et~al.}(2017){Oey}, {Herrera}, {Silich}, {Reiter}, {James},
  {Jaskot}, \& {Micheva}}]{oey17}
{Oey}, M.~S., {Herrera}, C.~N., {Silich}, S., {et~al.} 2017, ArXiv e-prints,
  arXiv:1710.03261

\bibitem[{{Robertson} {et~al.}(2015){Robertson}, {Ellis}, {Furlanetto}, \&
  {Dunlop}}]{robertson15}
{Robertson}, B.~E., {Ellis}, R.~S., {Furlanetto}, S.~R., \& {Dunlop}, J.~S.
  2015, \apjl, 802, L19

\bibitem[{{Roy} {et~al.}(1992){Roy}, {Aube}, {McCall}, \& {Dufour}}]{roy92}
{Roy}, J.-R., {Aube}, M., {McCall}, M.~L., \& {Dufour}, R.~J. 1992, \apj, 386,
  498

\bibitem[{{Schaerer} {et~al.}(2016){Schaerer}, {Izotov}, {Verhamme},
  {Orlitov{\'a}}, {Thuan}, {Worseck}, \& {Guseva}}]{schaerer16}
{Schaerer}, D., {Izotov}, Y.~I., {Verhamme}, A., {et~al.} 2016, \aap, 591, L8

\bibitem[{{Schlafly} \& {Finkbeiner}(2011)}]{schlafly11}
{Schlafly}, E.~F., \& {Finkbeiner}, D.~P. 2011, \apj, 737, 103

\bibitem[{{Shapley} {et~al.}(2016){Shapley}, {Steidel}, {Strom},
  {Bogosavljevi{\'c}}, {Reddy}, {Siana}, {Mostardi}, \& {Rudie}}]{shapley16}
{Shapley}, A.~E., {Steidel}, C.~C., {Strom}, A.~L., {et~al.} 2016, \apjl, 826,
  L24

\bibitem[{{Silich} \& {Tenorio-Tagle}(2017)}]{silich17}
{Silich}, S., \& {Tenorio-Tagle}, G. 2017, \mnras, 465, 1375

\bibitem[{{Silich} {et~al.}(2007){Silich}, {Tenorio-Tagle}, \&
  {Mu{\~n}oz-Tu{\~n}{\'o}n}}]{silich07}
{Silich}, S., {Tenorio-Tagle}, G., \& {Mu{\~n}oz-Tu{\~n}{\'o}n}, C. 2007, \apj,
  669, 952

\bibitem[{{Smith} {et~al.}(2006){Smith}, {Westmoquette}, {Gallagher},
  {O'Connell}, {Rosario}, \& {de Grijs}}]{smith06}
{Smith}, L.~J., {Westmoquette}, M.~S., {Gallagher}, J.~S., {et~al.} 2006,
  \mnras, 370, 513

\bibitem[{{Stanway} {et~al.}(2016){Stanway}, {Eldridge}, \&
  {Becker}}]{stanway16}
{Stanway}, E.~R., {Eldridge}, J.~J., \& {Becker}, G.~D. 2016, \mnras, 456, 485

\bibitem[{{Tenorio-Tagle} {et~al.}(2010){Tenorio-Tagle}, {W{\"u}nsch}, {Silich},
 {Mu{\~n}oz-Tu{\~n}{\'o}n}, \& {Palou{\v s}}}]{tenorio10}
{Tenorio-Tagle}, G., {W{\"u}nsch}, R., {Silich}, S., {Mu{\~n}oz-Tu{\~n}{\'o}n}, C.,
 \& {Palou{\v s}}, J. 2010, \apj, 708, 1621

\bibitem[{{Trebitsch} {et~al.}(2017){Trebitsch}, {Blaizot}, {Rosdahl},
  {Devriendt}, \& {Slyz}}]{trebitsch17}
{Trebitsch}, M., {Blaizot}, J., {Rosdahl}, J., {Devriendt}, J., \& {Slyz}, A.
  2017, \mnras, 470, 224

\bibitem[{{Vanzella} {et~al.}(2016){Vanzella}, {de Barros}, {Vasei}, {Alavi},
  {Giavalisco}, {Siana}, {Grazian}, {Hasinger}, {Suh}, {Cappelluti}, {Vito},
  {Amorin}, {Balestra}, {Brusa}, {Calura}, {Castellano}, {Comastri}, {Fontana},
  {Gilli}, {Mignoli}, {Pentericci}, {Vignali}, \& {Zamorani}}]{vanzella16}
{Vanzella}, E., {de Barros}, S., {Vasei}, K., {et~al.} 2016, \apj, 825, 41

\bibitem[{{Verhamme} {et~al.}(2015){Verhamme}, {Orlitov{\'a}}, {Schaerer}, \&
  {Hayes}}]{verhamme15}
{Verhamme}, A., {Orlitov{\'a}}, I., {Schaerer}, D., \& {Hayes}, M. 2015, \aap,
  578, A7

\bibitem[{{Verhamme} {et~al.}(2017){Verhamme}, {Orlitov{\'a}}, {Schaerer},
  {Izotov}, {Worseck}, {Thuan}, \& {Guseva}}]{verhamme17}
{Verhamme}, A., {Orlitov{\'a}}, I., {Schaerer}, D., {et~al.} 2017, \aap, 597,
  A13

\bibitem[{{Westmoquette} {et~al.}(2014){Westmoquette}, {Bastian}, {Smith},
  {Seth}, {Gallagher}, {O'Connell}, {Ryon}, {Silich}, {Mayya},
  {Mu{\~n}oz-Tu{\~n}{\'o}n}, \& {Rosa Gonz{\'a}lez}}]{westmoquette14}
{Westmoquette}, M.~S., {Bastian}, N., {Smith}, L.~J., {et~al.} 2014, \apj, 789,
  94

\bibitem[{{Wise} \& {Cen}(2009)}]{wise09}
{Wise}, J.~H., \& {Cen}, R. 2009, \apj, 693, 984

\bibitem[{{Yadav} {et~al.}(2017){Yadav}, {Mukherjee}, {Sharma}, \&
  {Nath}}]{yadav17}
{Yadav}, N., {Mukherjee}, D., {Sharma}, P., \& {Nath}, B.~B. 2017, \mnras, 465,
  1720

\bibitem[{{Yajima} {et~al.}(2014){Yajima}, {Li}, {Zhu}, {Abel}, {Gronwall}, \&
  {Ciardullo}}]{yajima14}
{Yajima}, H., {Li}, Y., {Zhu}, Q., {et~al.} 2014, \mnras, 440, 776

\end{thebibliography}

\end{document}